\documentclass[prl,showpacs,twocolumn]{revtex4}
\usepackage{graphicx,pdfsync}
\newcommand{\be}{\begin{equation}}

\newcommand{\ee}{\end{equation}}
\newcommand{\bea}{\begin{eqnarray}}
\newcommand{\eea}{\end{eqnarray}}
\newcommand{\bse}{\begin{subequations}}
\newcommand{\ese}{\end{subequations}}

\begin{document}

\def\K{{{K}}}
\def\Q{{{Q}}}
\def\Gbar{\bar{G}}
\def\tk{\tilde{\bf{k}}}
\def\k{{{k}}}
\def\q{{\bf{q}}}

\title{Coexistence of strong nematic and superconducting correlations in 
  a two-dimensional Hubbard model}

\author{Shi-Quan Su$^{1,2,*}$, and Thomas A. Maier$^{1}$} 
\affiliation{
$^{1}$Computer Science and Mathematics Division, Center for Nanophase Materials
Sciences, Oak Ridge National Laboratory, Oak Ridge, Tennessee 37831-6164, USA \\
$^{2}$National Center for Computational Sciences, Oak Ridge National Laboratory,
Oak Ridge, Tennessee 37831-6164, USA \\ $^{*}$ shiquansu@hotmail.com}

\begin{abstract}

Using a dynamic cluster quantum Monte Carlo approximation, we study a 
two-dimensional Hubbard model with a small orthorhombic distortion in 
the nearest neighbor hopping integrals. We find a large nematic response 
in the low-frequency single-particle scattering rate which develops with 
decreasing temperature and doping as the pseudogap region is entered.  
At the same time, the d-wave superconducting gap function develops an 
s-wave component and its amplitude becomes anisotropic. The strength of 
the pairing correlations, however, is found to be unaffected by the 
strong anisotropy, indicating that d-wave superconductivity can coexist 
with strong nematicity in the system.

\end{abstract} 

\maketitle


%
\paragraph*{Introduction-}
Nematic correlations, where the C$_4$ rotational symmetry of the square 
lattice is spontaneously broken and reduced to C$_2$, have been observed 
in the electronic structure in a number of strongly correlated electron 
materials, including the strontium-ruthenate oxide compounds, the 
cuprates and the iron-pnictide materials ~\cite{Fradkin10}. Despite the presence of very 
small structural anisotropies in the plane, these systems display very 
large nematic anisotropies in properties such as the dc and infrared 
resistivities ~\cite{Ando02}, the magnetoresistance ~\cite{Borzi}, the 
Nernst signal ~\cite{Daou}, the neutron scattering spectrum 
~\cite{Hinkov08} as well as in scanning tunneling spectroscopy 
~\cite{Kivelson,Lawler}. 

From a theoretical perspective, the idea that the pseudogap in the
cuprates may be associated with a nematic phase, which arises from
melting of a stripe ordered phase, was first discussed in
Ref.~\cite{Emery97} and later expanded on in Ref.~\cite{Kivelson98}. A
first microscopic theory of how a nematic phase can emerge from an
isotropic Fermi liquid through a Fermi surface instability was presented
in Ref.~\cite{Oganesyan01}. And indeed, both strong-coupling RVB
slave-boson mean-field calculations for the 2D t-J model
\cite{Yamase2000,Yamase2006} and weak-coupling functional
renormalization group studies \cite{halboth2000,grote2002,neumayr2003}
of the 2D Hubbard model have found a Pomeranchuk instability towards a
d-wave Fermi surface deformation (for a review see Ref.~\cite{vojta}).
This nematic instability has been found to compete with d-wave
superconductivity, a picture that has also emerged from exact
diagonalization \cite{miyanaga06} and variational Monte Carlo studies
\cite{edegger06} of the t-J model.

Here we are interested in the question of how a small orthorhombic 
distortion that exists in the system affects the electronic structure in 
a strongly correlated electron system. Consistent with experiments, 
recent theoretical studies have found a large nematic response to a 
small structural anisotropy. In particular, a recent study of the 2D 
Hubbard model using cellular dynamical mean-field theory and a dynamic 
cluster approximation (DCA) of a small 2$\times$2-site cluster has found 
a large nematic response in the low-frequency behavior of the 
single-particle self-energy as well as in the conductivity when the 
system is doped towards the Mott insulating state \cite{Okamoto}. This 
study was restricted to zero temperature and a small 2$\times$2 cluster.  
Here, using a larger 4$\times$4 cluster DCA calculation of a similar 2D 
Hubbard model with a small anisotropy in the single-particle hopping 
integral, we study the temperature dependence of the nematic response as 
well as its connection with the pseudogap and its interplay with the 
superconducting behavior of this model.

We find a large nematic anisotropy in the single-particle scattering
rate, which increases with decreasing temperature and doping.  We show
that this nematic anisotropy becomes significant inside the pseudogap
region below a temperature $T^*(\delta)$, pointing to a deep
relationship between the pseudogap and nematic correlations.  Moreover,
we find that the orthorhombicity in the single-particle hopping leads to
an anisotropic superconducting gap, with both $d$-wave and $s$-wave
components similar to what was discussed in Ref.~\cite{Emery97}, but has
no effect on the superconducting transition temperature.

\paragraph*{Formalism-}

The orthorhombic 2D Hubbard model we analyze \cite{Okamoto},
\begin{equation}
H = \sum_{\langle ij\rangle,\sigma} 
(t_{ij}-\mu\delta_{ij})c^\dagger_{i\sigma}c^{\phantom\dagger}_{j\sigma}+U\sum_{i}n_{{i}\uparrow}n_{{i}\downarrow} 
\label{eq:hubbard}\end{equation}
has a nearest neighbor transfer integral $t_{ij}=-t_{x/y}$ for hopping 
along the $x$- and $y$-direction, respectively, a Coulomb repulsion $U$ 
and chemical potential $\mu$. As in Ref.~\cite{Okamoto}, we use a small 
anisotropy $t_{x/y} = t(1\pm\xi/2)$ with $\xi=0.04$ to break 
the C$_4$ lattice symmetry of the model. In the following we will 
measure energies in units of $t$ and set the Coulomb repulsion $U=6t$
unless otherwise noted.

We will use the DCA to study the model in Eq.~(\ref{eq:hubbard}). The 
DCA ~\cite{hettler:dca, Maier05} maps the bulk lattice problem onto an 
effective periodic cluster embedded in a self-consistent dynamic 
mean-field that is designed to represent the remaining degrees of 
freedom. The DCA calculations were carried out on a 4$\times$4 cluster 
and the effective cluster problem was solved using a Hirsch-Fye quantum 
Monte Carlo algorithm \cite{jarrellQMC01}.

\paragraph*{Results-} In contrast to the 2$\times$2-site cluster study performed
in ~\cite{Okamoto}, we have found that, in the 4$\times$4-site cluster,
an additional next-nearest neighbor hopping $t'$ reduces the nematic
response relative to that found for $t'=0$. We therefore set $t'=0$ in
the 4$\times$4-site cluster calculations.

In order to examine how the directional anisotropy in the hopping
integral affects the Fermi surface of the correlated system, we plot in
Fig.~1 the gradient of the momentum distribution function, $|\nabla
n({\bf k})|$ with $n({\bf k}) = 2\langle c^\dagger_{{\bf k}\sigma}
c^{\phantom\dagger}_{{\bf k}\sigma}\rangle$, in the first quadrant of
the Brillouin zone. We have interpolated the DCA cluster self-energy
$\Sigma({\bf K},i\omega_n)$, which is defined on a 4$\times$4 grid of
${\bf K}$-points using a smooth spline. 

\begin{figure}[t] \includegraphics*[width=8cm]{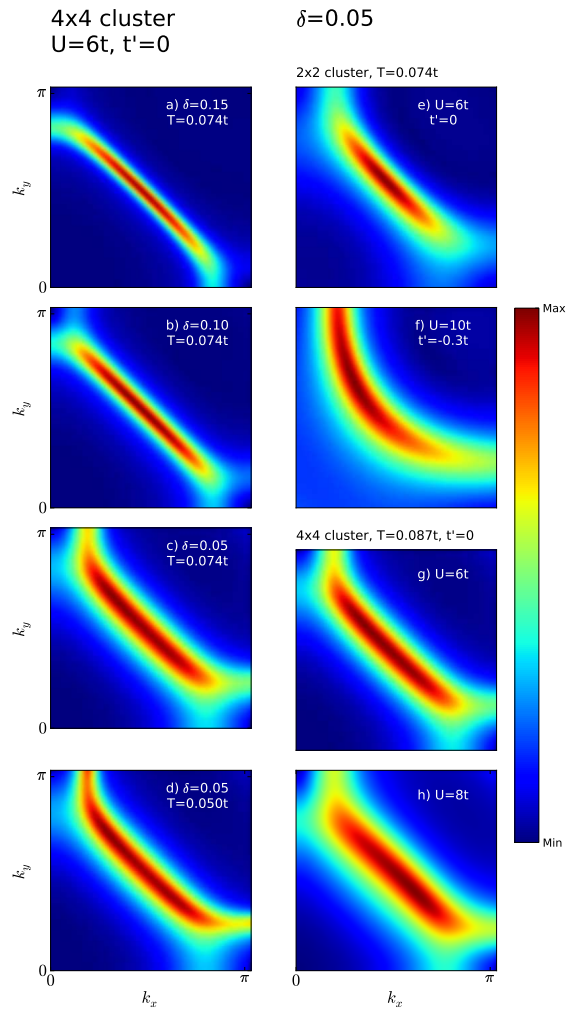}
  \caption{(Color online) Plots of the gradient of the k-space
    occupation, $|\nabla n_k|$ in the first quadrant of the Brillouin
    zone, calculated for the system with small directional anisotropy.
    On the left, results are shown for the 4$\times$4 cluster with
    $U=6t$ and $t'=0$ in panels a) through c) for a fixed temperature
    $T=0.074t$ but varying doping as indicated, and in panel d) for a
    lower temperature $T=0.05t$ and a doping $\delta=0.05$. The right
    side illustrates the effects of cluster size and varying parameters
    for a doping $\delta=0.05$.  Panels e) and f) show results for a
    2$\times$2 cluster for different values of $U$ and $t'$ as
    indicated. Panels g) and f) show results for the 4$\times$4 cluster
    for $t'=0$ for different values of $U$.}
\label{fig1} 
\end{figure}

The panels on the left show the results for the 4$\times$4-site cluster
for $U=6t$ and $t'=0$. Panels a) through c) show results for different
dopings $\delta$ calculated at a fixed temperature $T=0.074t$, while
panel d) shows the result for a lower temperature $T=0.05t$ and a doping
of $\delta=0.05$. From this one sees that at high temperatures and
doping levels, $|\nabla n({\bf k})|$ is fairly symmetric. But at low
temperatures and dopings, directional anisotropy clearly develops in
both its shape and, much more noticable, in its magnitude, with a larger
value in the region near $(0,\pi)$ than near $(\pi,0)$. As pointed out
in Ref.~\cite{Okamoto}, this reflects a stronger anisotropy in the
imaginary part of the self-energy than in the real part.

This result is supported by the plots of the real and imaginary parts of
the self-energy in Fig.~2. There one clearly sees that the imaginary
part exhibits qualitatively different behavior between $(\pi,0)$ and
$(0,\pi)$, while the real part shows much less anisotropic. For ${\bf
K}=(0,\pi)$, $|{\rm Im}\,\Sigma({\bf K},\omega_n)|$ displays Fermi
liquid like behavior with a downturn at low frequencies, while for ${\bf
K}=(\pi,0)$, it keeps rising with decreasing $\omega_n$. A similarly
strong angle dependence of the self-energy in the nematic phase was also
found in Ref.~\cite{Oganesyan01}. In addition, one sees that the
anisotropy in the imaginary part goes away at high frequencies,
consistent with the behavior found in Ref.~\cite{Okamoto}.

\begin{figure}[t]
\includegraphics*[width=8.5cm]{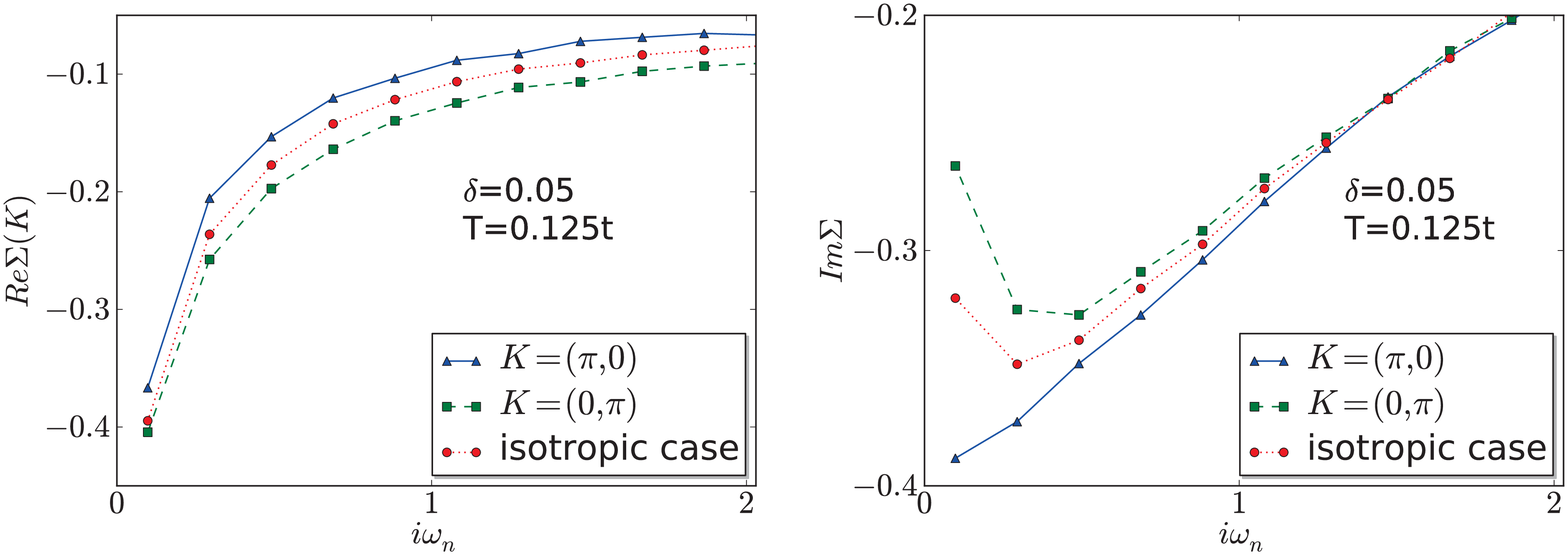}
\caption{(Color online)  Real and imaginary parts of the self-energy 
  $\Sigma({\bf K},\omega_n)$ versus frequency $\omega_n$ for ${\bf 
    K}=(\pi,0)$ and $(0,\pi)$ for a doping $\delta=0.05$ and temperature 
  $T=0.125t$. Results for both the anisotropic and isotropic cases are 
  shown.} \label{fig2}
\end{figure}

We generally find that the nematic response in the 4$\times$4-site
cluster is not larger than that found in the 2$\times$2-site cluster. As
an example, we show on the right hand side of Fig.~1 in panels e) and f)
results for the 2$\times$2-site cluster for $U=6t$ and $t'=0$ and
$U=10t$ and $t'=-0.3t$, respectively. Here one sees that in the 4-site
cluster, the nematic response is dramatically increased for larger
values of $U$ and a finite $t'=-0.3t$. In the 16-site cluster, however,
this does not seem to be the case. While a systematic study of larger
values of $U$ and finite $t'$ is not possible in the 16-site cluster,
because of the QMC sign problem, we show as an example in panels g) and
h) 16-site cluster results for two different values of $U$, $U=6t$ and
$8t$, respectively. There is no obvious enhancement of the response for
the larger value of $U$.

In order to further examine the temperature and doping dependence of the
nematic response, we have calculated the quasiparticle weight
approximated by
\begin{equation}
  Z_{\bf K} = \left[ 1-\mathrm{Im} \frac{\Sigma({\bf K},\pi T)}{\pi 
      T}\right]^{-1}.
\end{equation}
In Fig.~3 we plot the nematic anisotropy of $Z_{\bf K}$ defined as
\begin{equation}
  \sigma_Z  = \frac{Z_{(0,\pi)}-Z_{(\pi,0)}}{Z_{(0,\pi)}+Z_{(\pi,0)}}\,
\end{equation}
for the 4$\times$4-site cluster. One clearly sees that the nematic
response increases with decreasing temperature and doping, reaching
remarkable levels of over $50\%$ at the lowest temperature and doping
level we have studied.

Also indicated in this plot by the solid blue line is the pseudogap
temperature $T^*(\delta)$ as in ~\cite{Yang11}. Below this temperature,
a pseudogap opens in the single-particle density of states as well as in
the spin excitations. Here we have determined $T^*(\delta)$ from the
temperature $T$ at which the uniform magnetic susceptibility
$\chi_s({\bf q}=0,T)$ has a maximum ~\cite{Su}. It is interesting to note that this
temperature is not affected by the orthorhombic anisotropy in the
Hamiltonian. Conversely, as evidenced by this plot, the emergence of a
strong nematic response is intimately linked to the opening of the
pseudogap. Outside the pseudogap region, the nematic anisotropy is
small. With decreasing temperature and doping, it increases dramatically
along the pseudogap temperature $T^*(\delta)$ and then saturates deep
inside the pseudogap region. This points towards the pseudogap being a
necessary precondition for nematic physics to appear in this system.

\begin{figure}[t]
\includegraphics*[width=8cm]{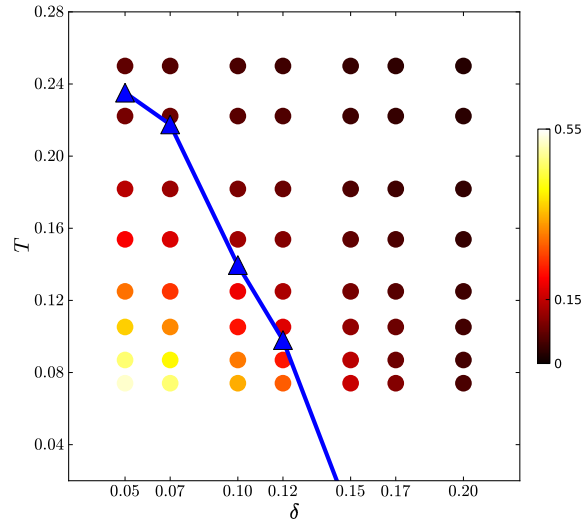}
\caption{ (Color online) Color map of the anisotropic ratio of the 
  quasiparticle weight, $ \sigma_Z $, over the temperature-doping plane, 
  for U=6t. The blue curve indicates the pseudogap temperature 
  $T^*(\delta)$ which is obtained as the temperature at which
  the uniform magnetic susceptibility $\chi_m(q=(0,0),T)$ has a 
  maximum.} \label{fig3}
\end{figure}

Finally, we note that we do not find spontaneous symmetry breaking to a nematic
state in the C$_4$ symmetric model, i.e.  for $\xi=0$, for the
temperatures we have studied. Note, however, that this does not exclude
the possibility of an instability at a lower temperature. In fact, the
large nematic response we find at the temperatures we have studied is
strongly suggestive of an instability at lower temperatures. A detailed
study of this issue is beyond the scope of the present paper and will
be the subject of future work.

It is interesting to ask how the strong nematic anisotropy that is
present in the single-particle scattering rate affects the
superconducting behavior of the system. In Ref.~\cite{Emery97} it was
pointed out that, in a nematic phase, the superconducting order
parameter will have mixed $d_{x^2-y^2}$ and extended $s$ symmetry. This
question was also addressed in a previous slave-boson mean-field study
of an anisotropic t-J model in Ref.~\cite{Yamase2000}.

Here we study the superconducting behavior of the anisotropic Hubbard 
model (\ref{eq:hubbard}) by solving the Bethe-Salpeter equation in the 
particle-particle channel
\begin{equation}
-\frac{T}{N}\sum_{K'}\Gamma^{pp}(K,K')\chi_0^{pp}(K')\phi_\alpha(K')=\lambda_\alpha 
\phi_\alpha(K) \,,
\label{eq:pairingmatrix}
\end{equation}
where $K=({\bf K},\omega_n)$ and $\Gamma^{pp}(K,K')$ is the irreducible 
particle-particle vertex with center of mass momentum and frequency 
$Q=0$. The coarse-grained bare particle-particle Green's function 
$\chi_0^{pp}(K) = N_c/N\sum_{\tilde{\bf k}}G_\uparrow({\bf K}'+{\bf 
  k}',\omega_n) G_\downarrow(-{\bf K}-{\bf k},-\omega_n)$ is calculated 
from the lattice Green's function $G_\sigma({\bf 
  k},\omega_n)=[i\omega_n-\epsilon_{{\bf k}}+\mu-\Sigma({\bf 
  K},\omega_n)]^{-1}$. The system undergoes a superconducting transition 
at the temperature $T_c$ where the leading eigenvalue $\lambda_\alpha$ 
becomes one, and the momentum and frequency dependence of the 
corresponding eigenvector $\phi_\alpha(K)$ determines the symmetry of 
the superconducting state. At $T_c$, $\phi_\alpha(K)$ becomes identical 
to the superconducting gap function. In the isotropic model, the 
instability occurs in the spin singlet, even frequency channel and the 
leading eigenvector has $d$-wave symmetry \cite{maier06}.  Here we focus 
on the low-doping region, where the nematic response of the anisotropic 
model is most pronounced. 

\begin{figure}[t] \includegraphics*[width=9.0cm]{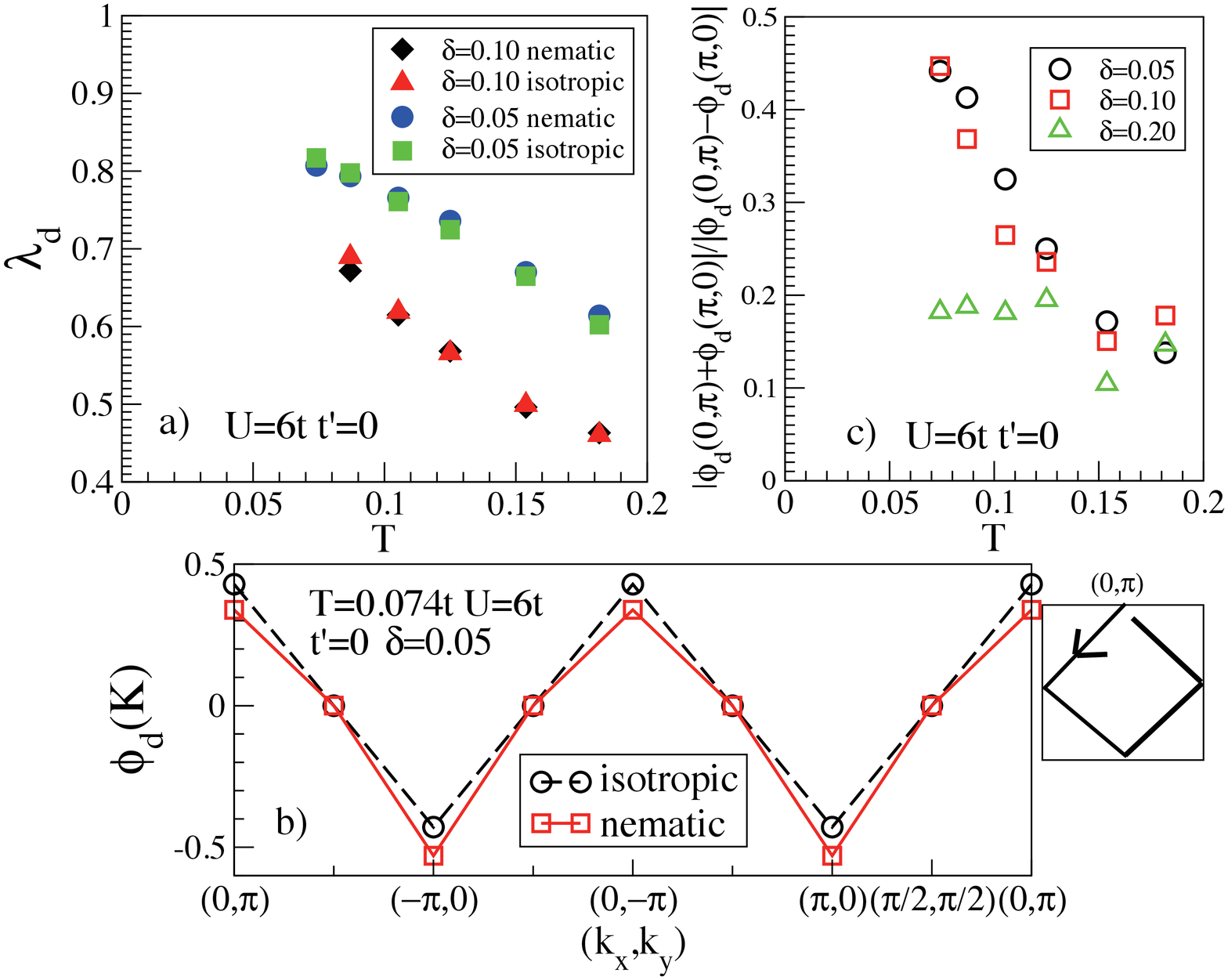}
  \caption{(Color online) (a) Leading d-wave eigenvalues $\lambda_d$ of
    the Bethe-Salpeter equation~(\ref{eq:pairingmatrix}) in the
    particle-particle channel versus temperature. The eigenvalues for
    both the isotropic and anisotropic models are plotted.  (b) The
    momentum dependence of the leading eigenvector $\phi_d(K,\pi T)$ at
    the lowest Matsubara frequency along the path in Brillouin zone
    shown in the inset.  (c) The relative s-wave contribution to the
    d-wave contribution,
    $|(\Phi_d(\pi,0)+\Phi_d(0,\pi))/(\Phi_d(\pi,0)-\Phi_d(0,\pi))|$, as
    a function of temperature for different dopings. } \label{fig4}
\end{figure}

Fig.~4a shows the temperature dependence of the leading eigenvalues of 
the anisotropic and isotropic models for two different dopings 
$\delta=0.10$ and $0.05$. In the temperature range we could access --- 
the QMC Fermion sign problem prevents us from going to lower 
temperatures --- the leading eigenvalues of the isotropic ($\xi=0$) 
and anisotropic ($\xi=0.04$) models are essentially identical. This 
indicates that the strong nematic anisotropy which is present in the 
single-particle scattering rate has essentially no effect on the 
strength of the pairing correlations. We also calculated the leading 
eigenvalue in an 8-site cluster, which allows us to reach the 
temperature where the eigenvalue crosses one. There we find that the 
superconducting transition temperature $T_c$ of the anisotropic model is 
unchanged from the isotropic case. 

In Fig.~4b we plot the momentum dependence of the leading eigenvector 
$\phi({\bf K},\pi T)$ along the line indicated in the inset. From this 
one sees that in the anisotropic case, it has predominantly d-wave 
structure, just as in the isotropic model. There is an additional s-wave 
contribution in the anisotropic model, which causes an anisotropy 
between $(\pi,0)$ and $(0,\pi)$, but leaves the nodal point unchanged. 

In Fig. ~4c, we plot the s-wave contribution related to d-wave $|\phi((\pi,0),\pi T)+\phi((0,\pi),\pi T)|/|\phi((\pi,0),\pi T)-\phi((0,\pi),\pi T)|$ as a function of temperature for different 
dopings. Once again, we find that the anisotropy in the eigenvector 
develops and increases when the system is cooled and doped into the 
pseudogap region. 



These results demonstrate that superconductivity can coexist with 
nematicity in the system. In addition, they indicate that the strength 
of the pairing correlations and $T_c$ are essentially unaffected by the 
large anisotropy present in the single-particle scattering rate. This is 
similar to what was found in earlier slave-boson mean-field calculations 
~\cite{Yamase2000}.

\paragraph*{Conclusions}
We have used a dynamic cluster quantum Monte Carlo approximation to study a two-dimensional Hubbard model with a small ($4\%$) orthorhombic distortion in the nearest-neighbor hopping integrals $t_x\neq t_y$. We have found a large nematic response, i.e. a difference in the electronic properties between ${\bf K}=(\pi,0)$ and $(0,\pi)$, with levels up to $55\%$ in the scattering rate ${\rm Im}\,\Sigma({\bf K},\omega_n)$, which develops as one enters the pseudogap phase and increases as the doping and temperature is further reduced. Similar behavior is found in the d-wave superconducting gap function, which develops an s-wave component in the pseudogap phase which reaches levels up to $40\%$. In contrast, we find that the strength of the pairing correlations is essentially unaffected by these anisotropies. These results demonstrate a close link between the pseudogap and nematic correlations, and indicate that d-wave superconductivity can coexist with a strong nematicity in the system.

\paragraph*{Acknowledgments-} We would like to thank S. Okamoto and
D.J. Scalapino for helpful discussion. Shi-Quan Su performed the
above research under contract number DE-AC05-00OR22750 between the
U.S. Department of Energy and Oak Ridge Associated Universities. A 
portion of this research was conducted at the Center for Nanophase 
Materials Sciences, which is sponsored at Oak Ridge National Laboratory 
by the Office of Basic Energy Sciences, U.S. Department of Energy.  This 
research used resources of the Oak Ridge Leadership Computing
Facility at Oak Ridge National Laboratory, which is supported by the
Office of Science of the U.S. Department of Energy under Contract No.
DE-AC05-00OR22725.


\begin{thebibliography}{99}

\bibitem{Fradkin10} E. Fradkin, et al., Annu. Rev. Condens. Matter Phys. {\bf{1}}, 153 (2010).

\bibitem{Ando02} Y. Ando, K. Segawa, S. Komiya, A. N. Lavrov, Phys. Rev. 
Lett. {\bf{88}}, 137005 (2002).

\bibitem{Borzi} R. A. Borzi, et al., science, {\bf{315}}, 214 (2007).

\bibitem{Daou} R. Daou, et al., nature, {\bf{463}}, 519 (2010).

\bibitem{Hinkov08} V. Hinkov, et al., science, {\bf{319}}, 597 (2008).

\bibitem{Kivelson} S. A. Kivelson, et al. Rev. Mod. Phys. {\bf 75}, 1201 
(2003).

\bibitem{Lawler} M. J. Lawler, et al., Nature, {\bf{466}}, 347 (2010).

\bibitem{Emery97} V. Emery et al., Phys.\ Rev.\ B {\bf{56}}, 6120 (1997).

\bibitem{Kivelson98} S. A. Kivelson et al., Nature {\bf{393}}, 550 (1998).

\bibitem{Oganesyan01} V. Oganesyan et al., Phys.\ Rev.\ B {\bf{64}}, 195109, (2001).

\bibitem{Yamase2000} H. Yamase and H. Kohno, J. Phys. Soc. Jpn., {\bf 
  69}, 2151 (2000).

\bibitem{Yamase2006} H. Yamase and W. Metzner, Phys. Rev. B, {\bf 73}, 
214517 (2006).

\bibitem{halboth2000} C. J. Halboth and W. Metzner, Phys. Rev.  Lett., 
{\bf 85}, 5162 (2000).

\bibitem{grote2002} I. Grote, E. K\"ording, and F. Wegner,
{\bf 126}, 13854 (2002).

\bibitem{neumayr2003} A. Neumayr and W. Metzner, Phys. Rev. B, {\bf 67}
035112, (2003).

\bibitem{vojta} M. Vojta, Adv. in Phys., {\bf 58}, 699, (2009).

\bibitem{miyanaga06} A. Miyanaga and H. Yamase, Phys. Rev. B, {\bf 73} 
174513  (2006).

\bibitem{edegger06} B. Edegger, V. N. Muthukumar, and C. Gros, Phys.  
Rev. B, {\bf 74}, 165109 (2006).

\bibitem{hettler:dca} M.\ H.\ Hettler, et al., Phys.\ Rev.\ B {\bf{58}}, 
R7475 (1998); M.\ H.\ Hettler, et al., Phys.\ Rev.\ B {\bf{61}},12739 
(2000).

\bibitem{Maier05} T. Maier, M. Jarrell, T. Pruschke, and M. H. Hettler, 
Rev. Mod. Phys. {\bf 77}, 1027 (2005).

\bibitem{jarrellQMC01} M. Jarrell, T. Maier, C. Huscroft, and S.  
Moukouri, Phys. Rev. B, {\bf 64}, 195130, (2001).

\bibitem{Okamoto} S. Okamoto, D. Senechal, M. Civelli, and A. -M. S. 
Tremblay, Phys. Rev. B {\bf{82}}, 180511(R) (2010).


\bibitem{Yang11}
S. -X. Yang et al., Phys. Rev. Lett. {\bf 106}, 047004 (2011).

\bibitem{Su}
S. -Q. Su, et al., Phys. Rev. A {\bf{81}}, 051604(R) (2010), S. -Q. Su, et al., Phys. Rev. B {\bf{80}}, 104517 (2009).

\bibitem{maier06} T. Maier, M. Jarrell, and D. Scalapino, Phys.  Rev.  
Lett., {\bf 96}, 047005 (2006).
\end{thebibliography}
\end{document}